\documentstyle[12pt]{article}

\begin{document}
\baselineskip 0.7cm

\newcommand{\gsim}{ \mathop{}_{\textstyle \sim}^{\textstyle >} }
\newcommand{\lsim}{ \mathop{}_{\textstyle \sim}^{\textstyle <} }
\newcommand{\EV}{ {\rm eV} }
\newcommand{\KEV}{ {\rm keV} }
\newcommand{\MEV}{ {\rm MeV} }
\newcommand{\GEV}{ {\rm GeV} }
\newcommand{\TEV}{ {\rm TeV} }
\renewcommand{\thefootnote}{\fnsymbol{footnote}}
\setcounter{footnote}{1}

\begin{titlepage}
\begin{flushright}
UT-774
\\
May, 1997
\end{flushright}

\vskip 0.35cm
\begin{center}
{\large \bf 
Direct-Transmission Models of Dynamical Supersymmetry Breaking
}
\vskip 1.2cm
Izawa K.-I., Y.~Nomura, K.~Tobe,\footnote{Fellow of the Japan Society
for the Promotion of Science.} and T.~Yanagida
\vskip 0.4cm

{\it Department of Physics, University of Tokyo,\\
     Bunkyo-ku, Hongou, Tokyo 113, Japan}

\vskip 1.5cm

\abstract{We systematically construct gauge-mediated
supersymmetry(SUSY)-breaking models
with direct transmission of SUSY-breaking effects
to the standard-model sector. We obtain a natural model with the gravitino
mass $m_{3/2}$ smaller than $1~\KEV$ as required from the standard cosmology.
If all Yukawa coupling constants are of order one,
the SUSY-breaking scale $m_{SUSY}$
transmitted into the standard-model sector is given by $m_{SUSY} \simeq 
0.1 \frac{\alpha_i}{4 \pi} \Lambda$ where $\Lambda$ is the original dynamical 
SUSY-breaking scale. Imposing $m_{SUSY} \simeq (10^2-10^3)$ GeV, we get
$\Lambda \simeq (10^5-10^6)$ GeV, which yields the gravitino mass 
$m_{3/2}\simeq (10^{-2}-1)$ keV.}
\end{center}
\end{titlepage}

\renewcommand{\thefootnote}{\arabic{footnote}}
\setcounter{footnote}{0}

%
%
%
%

\section{Introduction}

Low-energy dynamical supersymmetry (SUSY) breaking with gauge mediation
is extremely attractive,
since it may not only solve various phenomenological problems
but also its dynamical nature may provide a 
natural explanation of the large hierarchy between the electroweak and some 
higher (say the Planck) scales \cite{Dine_review}. Several mechanisms
\cite{ADS,ISS,IY,Randall}
for dynamical SUSY breaking have been discovered 
and their applications to realistic models have been also proposed
\cite{DNS,HIY, MH}.

Structures of the proposed models
\cite{DNS,HIY, MH}
predict a relatively large
SUSY-breaking scale $\Lambda > 10^6~ \GEV$
to provide sufficiently large soft 
masses in the SUSY standard-model sector. 
On the other hand, the unclosure condition of our universe yields a constraint
on the gravitino mass as $m_{3/2} \lsim 1 ~\KEV$
\cite{gravitino_mass},
which corresponds
to the SUSY-breaking scale $\Lambda \lsim 10^6 ~\GEV$.
This is not achieved in the referred models.
In fact, a detailed analysis
\cite{MM}
on the models in Ref.~\cite{DNS}
has shown that the gravitino is likely to be heavier 
than $1~\KEV$,
which necessitates a late-time entropy production
\cite{MM,MMY} 
to dilute the gravitino energy density in the universe. 

In this paper, we systematically construct gauge-mediated models
of low-energy SUSY breaking
with the structure of direct transmission
(that is, without messenger gauge interactions).
We obtain models in which
the gravitino mass can be set smaller than $1~ \KEV$. 
The existence of such models suggests that low-energy dynamical SUSY breaking
with gauge mediation does not necessarily require 
complicated non-standard cosmology.

\section{Dynamical scale generation}

We first discuss a dynamics for scale generation since it is crucial
for the dynamical SUSY breaking in our models. We adopt a SUSY SU(2)
gauge theory with four doublet chiral superfields $Q_i$,
where $i$ is a flavor index ($i=1,\cdots,4$).
Without a superpotential, this theory has a flavor SU(4)$_F$ symmetry.
This SU(4)$_F$ symmetry is explicitly broken down to a global SP(4)$_F$
by a superpotential in our models.
We add gauge singlets $Y^a$ ($a=1, \cdots, 5$) which constitute
a five-dimensional representation of SP(4)$_F$
to obtain a tree-level superpotential
\begin{eqnarray}
 W_Y = \lambda_Y Y^a (QQ)_a,
\end{eqnarray}
where $(QQ)_a$ denote a five-dimensional representation
of SP(4)$_F$ given by a suitable combination of gauge invariants $Q_iQ_j$.

An effective superpotential
\cite{IS}
which describes the dynamics of the SU(2) gauge interaction
may be given by
\begin{eqnarray}
 W_{eff}=S(V^2 + V_a^2 - \Lambda^4) + \lambda_Y Y^a V_a
\label{dynamical_potential}
\end{eqnarray}
in terms of low-energy degrees of freedom
\begin{eqnarray}
 V \sim (QQ), \quad V_a \sim (QQ)_a,
\end{eqnarray}
where $S$ is an additional chiral superfield, $\Lambda$ is a dynamically
generated scale, and a gauge invariant ($QQ$) denotes a singlet of
SP(4)$_F$ defined by
\begin{eqnarray}
(QQ)=\frac{1}{2} (Q_1 Q_2 + Q_3 Q_4).
\end{eqnarray}
The effective superpotential Eq.(\ref{dynamical_potential}) implies that 
the singlet $V \sim (QQ)$
condenses as 
\begin{eqnarray}
\label{VEV}
 \langle V \rangle = \Lambda^2,
\end{eqnarray}
and SUSY is kept unbroken in this unique vacuum.
Since the vacuum preserves the flavor SP(4)$_F$ symmetry, we have no 
massless Nambu-Goldstone boson. The absence of flat direction at this stage
is crucial for causing dynamical SUSY breaking as seen in the next section.

\section{Dynamical SUSY breaking}

Let us further introduce a singlet chiral superfield $Z$
to consider a superpotential for dynamical SUSY breaking
\cite{IY}:
\begin{eqnarray}
 W_0 = W_Y + \lambda Z (QQ).
\end{eqnarray}

For a relatively large value of the coupling $\lambda_Y$,
we again obtain the condensation Eq.(\ref{VEV})
with the low-energy effective superpotential approximated by
\begin{eqnarray}
 W_{eff} \simeq \lambda \Lambda^2 Z.
\end{eqnarray}

On the other hand, 
the effective K\"ahler potential is expected to take a form
\begin{eqnarray}
 K = |Z|^2 - \frac{\eta}{4 \Lambda^2}|\lambda Z|^4 + \cdots,
\end{eqnarray}
where $\eta$ is a real constant of order one.

The effective potential for the scalar $Z$ (with the same notation as 
the superfield) is given by
\begin{eqnarray}
 V_{eff} \simeq |\lambda|^2 \Lambda^4 (1 +
         \frac{\eta}{\Lambda^2} |\lambda|^4 |Z|^2).
\end{eqnarray}
If $\eta > 0$, this implies $\langle Z \rangle = 0$.
Otherwise we expect $|\lambda \langle Z \rangle| \sim \Lambda$,
since the effective potential is
lifted in the large $|Z|$ ($> \Lambda$) region \cite{IY,HIY,Shirman}.
Anyway, the $F$-component of $Z$ superfield has nonvanishing
vacuum-expectation
value, $\langle F_Z \rangle \simeq \lambda \Lambda^2$, and thus SUSY is 
dynamically broken in this model.

In the following analyses, we assume the latter case
$|\lambda \langle Z \rangle| \sim \Lambda$,
which results in the breakdown of $R$ symmetry.%
\footnote{The spontaneous breakdown of the $R$ symmetry produces a
Nambu-Goldstone $R$-axion. This $R$-axion is, however, cosmologically
harmless, since it acquires a mass from the $R$-breaking constant term in the
superpotential which is necessary to set the cosmological constant to 
zero\cite{Bag}.
Modifications for the case $\langle Z \rangle = 0$
is touched upon in the final section.}

\section{One-singlet model}
\label{one_singlet}

Let us first consider a realistic model with one singlet
$Z$ for SUSY breaking which couples directly to $(QQ)$. It is referred as 
a `multiplier' singlet, hereafter.
We introduce four pairs of
massive chiral superfields $d$, $\bar{d}$, $l$, $\bar{l}$, $d'$, $\bar{d}'$,
and $l'$, $\bar{l'}$ which are all singlets under the strong SU(2).
We assume that the $d$, $d'$ and $\bar{d}$, $\bar{d}'$
transform as the down quark and its antiparticle, respectively, under the 
standard-model gauge group. The $l$, $l'$ and $\bar{l}$, $\bar{l}'$
are assumed to transform as the lepton doublet and its antiparticle, 
respectively. These fields are referred as messenger quarks and leptons.

The superpotential of the one-singlet model is given by
\begin{eqnarray}
 W_1 = W_Y + Z(\lambda (QQ) + k_d d {\bar d} + k_l l {\bar l})
     + m_d d {\bar d}'+ m_{\bar d} d' {\bar d} + m_l l {\bar l}'
     + m_{\bar l} l' {\bar l},
\end{eqnarray}
where m's denote mass parameters.%
\footnote{Dynamical generation of these mass terms will be discussed in 
the following sections.
Mass terms for SUSY-breaking transmission were considered
in Ref.\cite{HIY,Ran}.
In the course of writing this paper, we received a paper \cite{recent}
which also treated similar mass terms in SUSY-breaking models.}
For relatively small values of the couplings $k_d$ and $k_l$,
we have a SUSY-breaking vacuum with the vacuum-expectation values
of the messenger quarks and leptons vanishing.
Then the soft SUSY-breaking masses of the messenger quarks and leptons
are directly generated by $\langle F_Z \rangle =\lambda \Lambda^2\neq 0$ 
through the couplings $Z(k_d d {\bar d} + k_l l {\bar l})$.  

The above SUSY-breaking vacuum is the 
true vacuum as long as the mass parameters $m_\psi$ are much larger
than $\sqrt{k_\psi F_Z} \simeq \sqrt{k_\psi \lambda} \Lambda$
for $\psi=d,l$.
To find the stability condition of our vacuum, we examine the scalar potential
\begin{eqnarray}
  V&=&| \lambda \Lambda^2 + k_d d \bar{d}+ k_l l \bar{l} |^2
+ |m_d d|^2 + |m_l l|^2 + |m_{\bar{d}}\bar{d}|^2 + |m_{\bar{l}}\bar{l}|^2 
\nonumber \\
&& + |k_d Z \bar{d} + m_d \bar{d}'|^2 
+|k_d Z d + m_{\bar{d}} d'|^2
+ |k_l Z \bar{l} + m_l \bar{l}'|^2 
+|k_l Z l + m_{\bar{l}} l'|^2.
\end{eqnarray}
The vacuum
\begin{eqnarray}
\langle F_Z \rangle \simeq \lambda \Lambda^2, \quad
\langle d \rangle = \langle \bar{d} \rangle=\langle l \rangle
=\langle \bar{l} \rangle=\langle d' \rangle=\langle \bar{d}' \rangle
=\langle l' \rangle=\langle \bar{l}'\rangle=0
\end{eqnarray}
is stable when 
\begin{eqnarray}
|m_d m_{\bar{d}}|^2 &>& |k_d \langle F_Z \rangle|^2 ,
\nonumber \\
|m_l m_{\bar{l}}|^2 &>& |k_l \langle F_Z \rangle|^2.
\label{stable_cond}
\end{eqnarray}
In the following analysis, we restrict ourselves to
the parameter region Eq.(\ref{stable_cond}).

The standard-model gauginos acquire their masses through loops of the messenger
quarks and leptons when $\langle Z \rangle \neq 0$
(see Figs.\ref{gaugino_mass}-\ref{3F_gaugino_mass} and the Appendix). 
The gaugino masses are obtained as 
\begin{eqnarray}
m_{\tilde{g}_1} &=& \frac{\alpha_1}{4 \pi} \left\{
 \frac{2}{5} \left|\frac{k_d \langle F_Z \rangle}{m_d m_{\bar{d}}}
 \right|^2
 \frac{k_d \langle F_Z \rangle}{\sqrt{m_d m_{\bar{d}}}} 
 {\cal F}_d 
\right.
\nonumber \\
 &&\left. ~~+ \frac{3}{5} 
 \left|\frac{k_l \langle F_Z \rangle}{m_l m_{\bar{l}}}
 \right|^2
 \frac{k_l \langle F_Z \rangle}{\sqrt{m_l m_{\bar{l}}}}
 {\cal F}_l
\right\}\left(1+O((k_\psi \langle F_{Z}\rangle 
/m_\psi m_{\bar{\psi}})^2 \right),
\label{bino_mass}
\\
m_{\tilde{g}_2} &=& \frac{\alpha_2}{4 \pi}
\left| \frac{k_l \langle F_Z \rangle}{m_l m_{\bar{l}}} \right|^2
  \frac{k_l \langle F_Z \rangle}{\sqrt{m_l m_{\bar{l}}}}
  {\cal F}_l
\left(1+O((k_l\langle F_{Z}\rangle /m_l m_{\bar{l}})^2 \right),
\label{wino_mass}
\\
m_{\tilde{g}_3} &=& \frac{\alpha_3}{4 \pi}
\left| \frac{k_d \langle F_Z \rangle}{m_d m_{\bar{d}}} \right|^2
  \frac{k_d \langle F_Z \rangle}{\sqrt{m_d m_{\bar{d}}}}
  {\cal F}_d
\left(1+O((k_d \langle F_{Z}\rangle /m_d m_{\bar{d}})^2 \right),
\label{gluino_mass}
\end{eqnarray}
where we have adopted SU(5) GUT normalization of U(1)$_Y$ gauge coupling,
$\alpha_1 \equiv \frac{5}{3}\alpha_Y$, and $\tilde{g}_3, \tilde{g}_2$, and 
$\tilde{g}_1$ are gauginos of the
standard-model gauge groups SU(3)$_C$, SU(2)$_L$, and U(1)$_Y$, respectively. 
The ${\cal F}_\psi$ for $\psi=d,l$ are defined in the Appendix. 
Here, we have assumed 
$(k_\psi \langle F_Z \rangle/m_\psi m_{\bar{\psi}})^2 \ll 1$. 
Notice that
the leading term of $(k_\psi\langle F_Z \rangle/m_\psi m_{\bar{\psi}})$ 
in Fig.\ref{gaugino_mass} vanishes.
Hence the GUT relation among gaugino masses ,
$m_{\tilde{g}_1}/\alpha_1=m_{\tilde{g}_2}/\alpha_2=m_{\tilde{g}_3}/\alpha_3$,
does not hold even when
all the couplings and mass parameters for messenger quarks and 
leptons satisfy the GUT relation at the GUT scale.

The soft SUSY-breaking masses for squarks and sleptons $\tilde{f}$ in the
standard-model sector are generated
by two-loop diagrams shown in Fig.\ref{two_loop_sfermion_mass}.
We obtain them as
\begin{eqnarray}
m^2_{\tilde{f}}=2 
\left[ C_3^{\tilde{f}} \left(\frac{\alpha_3}{4 \pi} \right)^2
  \Lambda^{(d)2}
  + C_2^{\tilde{f}} \left(\frac{\alpha_2}{4 \pi} \right)^2
  \Lambda^{(l)2}
  + \frac{3}{5} Y^2 \left(\frac{\alpha_1}{4 \pi} \right)^2
  \left(\frac{2}{5} \Lambda^{(d)2}
  + \frac{3}{5} \Lambda^{(l)2} \right) \right],
\end{eqnarray}
where $C_3^{\tilde{f}}=\frac{4}{3}$ and $C_2^{\tilde{f}}=\frac{3}{4}$ 
when $\tilde{f}$ is in the fundamental representation of SU(3)$_C$ 
and SU(2)$_L$, and $C_i^{\tilde{f}}=0$ for the gauge singlets,
and $Y$ denotes the U(1)$_Y$ hypercharge
($Y \equiv Q-T_3$). Here the effective scales $\Lambda^{(\psi)}$
are of order $k_\psi \langle F_Z \rangle/m_\psi$. 
For example, the effective scales $\Lambda^{(\psi)}$
are given by
\begin{eqnarray}
\Lambda^{(\psi)2}=\frac{|k_\psi \langle F_Z \rangle|^2}{\bar{m}^2_\psi}
\end{eqnarray}
if the messenger quarks and leptons have a degenerate SUSY-invariant mass 
$\bar{m}_\psi$,%
\footnote{
In the present analysis, we only discuss
the sfermion masses qualitatively. A more detailed analysis
will be given in Ref.\cite{NT}.}
which is an eigenvalue of the mass matrix
\begin{eqnarray}
\left(
\begin{array}{cc}
k_\psi \langle Z \rangle & m_{\bar{\psi}}\\
m_{\psi} & 0
\end{array}
\right).
\end{eqnarray}

The SUSY-breaking squark 
and slepton masses are proportional to 
$(k_\psi \langle F_Z \rangle /m_\psi m_{\bar{\psi}})$.
On the other hand, the gaugino masses have an extra suppression
$(k_\psi \langle F_Z \rangle /m_\psi m_{\bar{\psi}})^2$ as shown in 
Eqs.(\ref{bino_mass})-(\ref{gluino_mass})
since the leading term of $(k_\psi\langle F_Z \rangle/m_\psi m_{\bar{\psi}})$ 
vanishes. Thus, to avoid 
too low masses for the gauginos, we must take 
$(k_\psi \langle F_Z \rangle /m_\psi m_{\bar{\psi}})^2 > 0.1$.
It is interesting that this condition is necessary to have a 
light gravitino with mass less than $1~\KEV$ as shown below.

We are now at a point to derive a constraint on the gravitino mass.
The conservative constraint comes from the experimental lower
bounds\footnote{
These bounds are derived assuming the GUT relation of the gaugino 
masses. The bound on the gluino mass assumes that the gluino is
heavier than all squarks. A more detailed phenomenological analysis on the 
models in this paper will be given in Ref.\cite{NT}.}
on the masses of wino and gluino\cite{LEP,PDG}\footnote{
We find in Ref.\cite{NT} that even when 
$(k \langle F_Z \rangle /m^2)^2 \simeq 1$, the
constraint from the right-handed slepton mass is weaker than those
from the gaugino masses.}
\begin{eqnarray}
m_{\tilde{g}_2} \gsim 50~\GEV,~~~m_{\tilde{g}_3} \gsim 220~\GEV,
\end{eqnarray}
which yield
\begin{eqnarray}
\left|\frac{k_l \langle F_Z \rangle}{m_l m_{\bar{l}}} \right|^2
\frac{k_l \langle F_Z \rangle}{\sqrt{m_l m_{\bar{l}}}} 
{\cal F}_l &\gsim& 1.9 \times10^4~\GEV,
\\
\left|\frac{k_d \langle F_Z \rangle}{m_d m_{\bar{d}}} \right|^2
\frac{k_d \langle F_Z \rangle}{\sqrt{m_d m_{\bar{d}}}} 
{\cal F}_d &\gsim& 2.3 \times10^4~\GEV.
\end{eqnarray}
We obtain 
\begin{eqnarray}
\langle F_Z \rangle &\gsim& \frac{3 \times 10^8}{k_l {\cal F}_l^2}
\left( \frac{m_l m_{\bar{l}}}{k_l \langle F_Z \rangle} \right)^5~\GEV^2,
\\
\langle F_Z \rangle &\gsim& \frac{5 \times 10^8}{k_d {\cal F}_d^2}
\left( \frac{m_d m_{\bar{d}}}{k_d \langle F_Z \rangle} \right)^5~\GEV^2.
\end{eqnarray}
The gravitino mass is given by 
\begin{eqnarray}
m_{3/2}&=&\frac{\langle F_Z \rangle}{\sqrt{3}M} \gsim
\frac{0.8}{k_l}\left(\frac{0.1}{{\cal F}_l}\right)^2
\left( \frac{m_l m_{\bar{l}}}{k_l \langle F_Z \rangle} \right)^5
\times 10^{-2}~ \KEV.
\\
m_{3/2}&=&\frac{\langle F_Z \rangle}{\sqrt{3}M} \gsim
\frac{1}{k_d}\left(\frac{0.1}{{\cal F}_d}\right)^2
\left( \frac{m_d m_{\bar{d}}}{k_d \langle F_Z \rangle} \right)^5
\times 10^{-2}~ \KEV.
\end{eqnarray}
Since the $|{\cal F}_\psi|$ has the maximal value $0.1$ 
(see the Appendix), we see that in the region of 
$0.2 \lsim (\frac{k_\psi \langle F_Z \rangle}{m_\psi m_{\bar{\psi}}})^2 
\lsim 1$ and $k_\psi \simeq 1$ for $\psi=d,l$,
the gravitino can be lighter than $1~\KEV$, which is required from the standard
cosmology.

We have found that the gravitino mass can be set smaller than 
$1~\KEV$ if $m_\psi$
are of order the SUSY-breaking scale $\Lambda$. In principle, the masses
$m_\psi$ of the messenger quarks and leptons might be considered 
to arise from dynamics of another strong 
interaction. In that case, however, it seems accidental to have 
$m_\psi \sim \Lambda$. Thus it is natural to consider a model in which the
SUSY-breaking dynamics produces simultaneously the mass terms for the
messenger quarks and leptons. This possibility will be 
discussed in section~\ref{three-singlet}.

We note that there is no CP violation in this model. All
the coupling constants $k_d,~k_l$ and the mass parameters $m_\psi$ 
($\psi=d,l,\bar{d},\bar{l}$) can be taken real without loss of 
generality. The vacuum-expectation values $\langle QQ \rangle$ and 
$\langle Z \rangle$ are also taken real by
phase rotations of the corresponding superfields.
Thus only the $\langle F_Z \rangle$ is a complex quantity
and then all the gaugino masses have
a common phase coming from the phase of $\langle F_Z \rangle$. However,
this phase can be eliminated by a common rotation
of the gauginos.\footnote{
The rotation of the gauginos induces a complex phase in the Yukawa-type
gauge couplings of the gauginos. However, such a complex phase is eliminated
by a rotation of the sfermions and Higgs fields $H$ and $\bar{H}$,
since we have no SUSY-breaking trilinear couplings and no SUSY-breaking
$B$ term $B\mu H {\bar{H}}$ at the tree-level.}

\section{Two-singlet model}

Next we consider a realistic model with two `multiplier' singlets
$Z_1$ and $Z_2$ for SUSY breaking.
We introduce two pairs of chiral superfields $d$, $\bar{d}$
and $l$, $\bar{l}$ which are all singlets under the strong SU(2).

We also introduce an additional singlet $X$ to obtain
a superpotential
\footnote{We could construct a model without the additional singlet
superfield \cite{HIY} at the sacrifice of complete naturalness. 
It may manage to accommodate a light gravitino with 
$m_{3/2}\sim 1~\KEV$ in a strong-coupling regime.}
\begin{eqnarray}
 W_2 = W_Y + Z_1(\lambda_1(QQ) - f_1X^2) + Z_2(\lambda_2(QQ) - f_2X^2)
     + X(f_dd{\bar d} + f_ll{\bar l}).
\end{eqnarray}
Without loss of generality, we may set $f_2 = 0$ by an appropriate redefinition
of $Z_1$ and $Z_2$.
Then the superpotential
yields a vacuum with $\langle X \rangle = \sqrt{f_1^{-1}\lambda_1} \Lambda$.
The masses of messenger quarks and leptons are given by 
\begin{eqnarray}
m_\psi=f_\psi \langle X \rangle
\end{eqnarray}
for $\psi=d,l$.
Since $F_{Z_2}=\lambda_2 \Lambda^2$ is nonvanishing, SUSY is broken.

The soft masses of the messenger quarks and leptons
stem from radiative corrections.
For example, the diagrams shown in Fig.\ref{Kahler_corr} generate an effective
K\"ahler potential of the form
\begin{eqnarray}
\frac{\delta}{(16 \pi^2)^2} |\lambda_1|^2 |\lambda_2|^2 
\lambda_1 |f_1|^2 f_1^*
\frac{Z_2^*Z_2(\lambda_1^* Z_1^* + \lambda_2^* Z_2^*) X^{*2} X}{\Lambda^6}
(|f_d|^2 f_d d{\bar d} + |f_l|^2 f_l l{\bar l}),
\end{eqnarray}
which gives soft mass terms of the form
\begin{eqnarray}
 \frac{\delta}{(16 \pi^2)^2}|\lambda_1|^2 |\lambda_2|^2 
\lambda_1 \lambda_2^*  |f_1|^2 f_1^*
\frac{|F_{Z_2}|^2 \langle Z_2 \rangle \langle X \rangle^3}{\Lambda^6}
 (|f_d|^2 f_d d{\bar d} + |f_l|^2 f_l l{\bar l}),
\end{eqnarray}
when $\langle Z_2 \rangle \neq 0$.

Since the induced soft masses for messenger squarks and sleptons are
suppressed by loop factors, the gravitino mass 
is expected to be much larger than $1~\KEV$ in this model.

\section{Three-singlet model}
\label{three-singlet}

We finally obtain a realistic model with three `multiplier' singlets
$Z_1$, $Z_2$, and $Z_3$ for SUSY breaking.
The model is a combination of the one- and the two-singlet models discussed
in the previous sections. 
The masses $m_\psi$ of messenger quarks and leptons in the one-singlet 
model are generated  by Yukawa couplings of X introduced in the two-singlet 
model.

The superpotential in this three-singlet model is given by
\begin{eqnarray}
 W_3 &=& W_Y + Z_1(\lambda_1(QQ) + k_{d1}d{\bar d} + k_{l1}l{\bar l} - f_1X^2)
     + Z_2(\lambda_2(QQ) + k_{d2}d{\bar d} + k_{l2}l{\bar l} - f_2X^2)
     \nonumber \\
     &&+ Z_3(\lambda_3(QQ) + k_{d3}d{\bar d} + k_{l3}l{\bar l} - f_3X^2)
     + X(f_d d {\bar d}' + f_{\bar d} d' {\bar d} + f_l l {\bar l}'
     + f_{\bar l} l' {\bar l}).
\label{superpotential_model3}
\end{eqnarray}
Without loss of generality, we may set
$k_{d1} = k_{l1} = f_2 = 0$
by an appropriate redefinition of $Z_1$, $Z_2$, and $Z_3$.
For relatively small values of the couplings
$k_{d2}$, $k_{l2}$, $\lambda_3$, $k_{d3}$, $k_{l3}$, and $f_3$,
the superpotential yields a vacuum with 
$\langle X \rangle = \sqrt{f_1^{-1}\lambda_1} \Lambda$
and the vacuum expectation values of the messenger quarks and leptons
vanishing.
The masses $m_\psi$ of messenger quarks and leptons in the one-singlet model
are given by 
\begin{eqnarray}
m_\psi=f_\psi \langle X \rangle
\end{eqnarray}
for $\psi=d,l,\bar{d}, \bar{l}$.
In this vacuum, the $F$-components of $Z_i$ are given by
\begin{eqnarray}
F_{Z_1} \simeq 0, \quad
F_{Z_2} \simeq \lambda_2 \Lambda^2, \quad
F_{Z_3} \simeq \lambda_3 \Lambda^2-f_3 \langle X \rangle^2,
\end{eqnarray}
and thus SUSY is broken.
The masses of gauginos, squarks, and sleptons are generated
as in the one-singlet model in section~\ref{one_singlet}. We should replace
$k_\psi \langle F_Z \rangle $ in Eqs.(\ref{bino_mass})-(\ref{gluino_mass}) by 
$k_{\psi 2} \langle F_{Z_2} \rangle+ k_{\psi 3} \langle F_{Z_3} \rangle$.

If $k_{d1}/k_{d2} \neq k_{l1}/k_{l2}$, the phases of the three
gauginos' masses are different from one another. Then, the phases of the 
gauginos' masses cannot be eliminated by a common rotation of the 
gaugino fields and thus CP is broken. However, there is no such problem
in the GUT models since $k_{d1}/k_{d2} \simeq k_{l1}/k_{l2}$ holds even at 
low energies.

We comment on the $\mu$-problem\cite{DNS,DGP}. If the superfield $X$ couples
to $H \bar{H}$ where $H$ and $\bar{H}$ are Higgs fields in the standard model,
the SUSY-invariant mass $\mu$ for Higgs $H$ and $\bar{H}$ is generated. To
have the desired mass $\mu \simeq (10^2-10^3)~\GEV$, we must choose a small
coupling constant $\lambda_h \simeq 10^{-3}$, where $\lambda_h$ is defined
by $W=\lambda_h X H \bar{H}$. This is natural in the sense of 't Hooft.
We note that no large $B$ term ($B\mu {H} {\bar{H}}$)
is induced since the $F$-component of $X$ is very small.
Hence the scale $\mu$ may originate from the SUSY-breaking scale
in the present model.%
\footnote{
There has been also proposed an interesting solution to the $\mu$-problem
in Ref.\cite{Yana}.}

Finally, we should stress that the superpotential
Eq.(\ref{superpotential_model3}) is natural, since it has a global symmetry
U(1)$_R \times$U(1)$_\chi$, where U(1)$_R$ is an $R$ symmetry. That is, the 
superpotential Eq.(\ref{superpotential_model3}) is a general one allowed
by the global U(1)$_R \times$U(1)$_\chi$.\footnote{
This global symmetry may forbid mixings between the messenger quarks and 
the down-type quarks in the standard-model sector. This avoids naturally
the flavor-changing neutral current problem\cite{DNS2}. Then
there exists the lightest stable particle in the messenger sector\cite{DGP2}.}
The charges for chiral superfields
are given in Table~\ref{table_charge}.
\section{Conclusion}

We have constructed gauge-mediated
SUSY-breaking models
with direct transmission of SUSY-breaking effects
to the standard-model sector. In our three-singlet model,
the gravitino mass $m_{3/2}$ is expected to be
smaller than $1~\KEV$ naturally as required from the standard cosmology:
If all the Yukawa coupling constants are of order one,
the SUSY-breaking scale $m_{SUSY}$
transmitted into the standard-model sector is given by $m_{SUSY} \simeq 
0.1 \frac{\alpha_i}{4 \pi} \Lambda$.
Imposing $m_{SUSY} \simeq (10^2-10^3)$ GeV, we get
$\Lambda \simeq (10^5-10^6)$ GeV, which yields the gravitino mass $m_{3/2}
\simeq (10^{-2}-1)$ keV.

In the present models, we have four gauge groups
SU(3)$_C \times$ SU(2)$_L \times$U(1)$_Y \times$SU(2). It is well known
that the three gauge coupling constants of the SUSY standard-model gauge groups
meet at the GUT scale $\sim 10^{16}~\GEV$. It is remarkable
that in the three-singlet model, all the four gauge coupling constants
meet at the scale $\sim 10^{16}~\GEV$ as shown in Fig.\ref{gauge_coupling}.
Here, we have assumed that the gauge coupling constant $\tilde{\alpha}_2$
of the strong SU(2) becomes strong ($\tilde{\alpha_2}/\pi \simeq 1$) at
the scale $\Lambda \simeq (10^5-10^6)~\GEV$.

So far we have assumed spontaneous breakdown of $R$ symmetry in the models.
If $\langle Z \rangle = 0$, we need to introduce $R$-breaking mass terms
such as $md{\bar d} + m'l{\bar l}$ to generate
the standard-model gaugino masses.
These mass terms might be induced through the $R$ symmetry breaking
which is necessary for the cosmological constant to be vanishing
\cite{Bag}.
%


\newpage

\section*{Appendix}

In this Appendix, we evaluate the standard-model gaugino masses in 
our SUSY-breaking models.
The superpotential which relates to the mass terms of messenger
fields $\psi$, $\bar{\psi}$, $\psi'$, and $\bar{\psi}'$ for $\psi=d,l$ is 
represented as
\begin{eqnarray}
W=\sum_{\psi=d,l}(\bar{\psi}, \bar{\psi}')
M^{(\psi)}
\left(
\begin{array}{c}
\psi \\
\psi'
\end{array}
\right),
\end{eqnarray}
where the mass matrix $M^{(\psi)}$ is given by
\begin{eqnarray}
M^{(\psi)}=
\left(
\begin{array}{cc}
m^{(\psi)}_1 & m^{(\psi)}_3\\
m^{(\psi)}_2 & 0
\end{array}
\right).
\label{mass_matrix}
\end{eqnarray}
In the one-singlet model, the mass parameters $m^{(\psi)}_i$
are given by 
\begin{eqnarray}
m^{(\psi)}_1 &=& k_\psi \langle Z \rangle,
\\
m^{(\psi)}_2 &=& m_\psi,
\\
m^{(\psi)}_3 &=& m_{\bar{\psi}},
\end{eqnarray}
and in the three-singlet model, they are given by
\begin{eqnarray}
m^{(\psi)}_1 &=& k_{\psi 2} \langle Z_2 \rangle 
+ k_{\psi 3} \langle Z_3 \rangle,
\\
m^{(\psi)}_2 &=& f_{\psi} \langle X \rangle ,
\\
m^{(\psi)}_3 &=& f_{\bar{\psi}} \langle X \rangle.
\end{eqnarray}

The soft SUSY-breaking mass terms of the messenger fields are given by
\begin{eqnarray}
{\cal{L}}_{soft}&=&\sum_{\psi=d,l} F^{(\psi)}\tilde{\psi} \tilde{\bar{\psi}},
\end{eqnarray}
where
\begin{eqnarray}
F^{(\psi)}=k_\psi \langle F_Z \rangle
\end{eqnarray}
in the one-singlet model and 
\begin{eqnarray}
F^{(\psi)}=k_{\psi 2} \langle F_{Z_2} \rangle 
+ k_{\psi 3} \langle F_{Z_3} \rangle
\end{eqnarray}
in the three-singlet model.
Then the standard-model gauginos acquire their masses through loops of the 
messenger
quarks and leptons. Their masses of order $F^{(\psi)}/m^{(\psi)}$
are given by (see Fig.\ref{gaugino_mass})
\begin{eqnarray}
m_{\tilde{g}_3} &=& \frac{\alpha_3}{4\pi} F^{(d)}
\left( M^{(d)^{-1}} \right)_{11},
\\
m_{\tilde{g}_2} &=& \frac{\alpha_2}{4\pi} F^{(l)}
\left( M^{(l)^{-1}} \right)_{11},
\\
m_{\tilde{g}_1} &=& \frac{\alpha_1}{4\pi} \left\{
\frac{2}{5} F^{(d)} \left( M^{(d)^{-1}} \right)_{11}
+\frac{3}{5} F^{(l)} \left( M^{(l)^{-1}} \right)_{11} 
\right\},
\end{eqnarray}
where the masses $m_{\tilde{g}_i}$ ($i=1, \cdots, 3$) denote bino, wino, and
gluino masses, respectively, and we have adopted the SU(5) GUT normalization
of the U(1)$_Y$ gauge coupling ($\alpha_1 \equiv \frac{5}{3} \alpha_Y$). 
Because of $\left( M^{(\psi)^{-1}} \right)_{11} =0$,
the above contributions vanish.
However, the contributions of higher powers of 
$F^{(\psi)}/m^{(\psi)2}$ do not vanish in general:
We now work in a basis where the supersymmetric masses $M^{(\psi)}$ are 
diagonalized as
\begin{eqnarray}
O_{\psi_\psi} M^{(\psi)} O_{\theta_\psi}^\dagger = 
\left(
\begin{array}{cc}
m_{\psi 1} & 0\\
0 & m_{\psi 2}
\end{array}
\right).
\end{eqnarray}
Here the mass eigenstates are given by
\begin{eqnarray}
\left(
\begin{array}{c}
\psi_1 \\
\psi_2
\end{array}
\right)
=O_{\theta_\psi}
\left(
\begin{array}{c}
\psi \\
\psi'
\end{array}
\right)
=
\left(
\begin{array}{cc}
\cos \theta_\psi & -\sin \theta_\psi \\
\sin \theta_\psi & \cos \theta_\psi 
\end{array}
\right)
\left(
\begin{array}{c}
\psi \\
\psi'
\end{array}
\right),
\\
\left(
\begin{array}{c}
\bar{\psi}_1 \\
\bar{\psi}_2
\end{array}
\right)
=O_{\phi_\psi}
\left(
\begin{array}{c}
\bar{\psi} \\
\bar{\psi}'
\end{array}
\right)
=
\left(
\begin{array}{cc}
\cos \phi_\psi & -\sin \phi_\psi \\
\sin \phi_\psi & \cos \phi_\psi 
\end{array}
\right)
\left(
\begin{array}{c}
\bar{\psi} \\
\bar{\psi}'
\end{array}
\right),
\end{eqnarray}
where we have taken the mass matrices $M^{(\psi)}$ to be real,
which is always possible.
%
%
Then, for example, the contribution of order 
$(F^{(\psi)}/m^{(\psi)})(F^{(\psi)}/m^{(\psi)2})^2$ to the gaugino
masses, which is shown in Fig.\ref{3F_gaugino_mass}, is represented by
\begin{eqnarray}
\label{g1_mass}
  m_{\tilde{g}_3}&=&\frac{\alpha_3}{4 \pi} \left|
  \frac{F^{(d)}}{m^{(d)}_2 m^{(d)}_3} \right|^2 \frac{F^{(d)}}
  {\sqrt{m^{(d)}_2 m^{(d)}_3}}
  {\cal F}_d,  \\
\label{g2_mass}
  m_{\tilde{g}_2}&=&\frac{\alpha_2}{4 \pi} \left|
  \frac{F^{(l)}}{m^{(l)}_2 m^{(l)}_3} \right|^2 \frac{F^{(l)}}
  {\sqrt{m^{(l)}_2 m^{(l)}_3}}
  {\cal F}_l, \\
\label{g3_mass}
  m_{\tilde{g}_1}&=&\frac{\alpha_1}{4 \pi} \left\{ \frac{2}{5}\left|
  \frac{F^{(d)}}{m^{(d)}_2 m^{(d)}_3 } \right|^2 \frac{F^{(d)}}
  {\sqrt{m^{(d)}_2 m^{(d)}_3}}
  {\cal F}_d
+ \frac{3}{5}\left|
  \frac{F^{(l)}}{m^{(l)}_2 m^{(l)}_3} \right|^2 \frac{F^{(l)}}
  {\sqrt{m^{(l)}_2 m^{(l)}_3}}
  {\cal F}_l \right\}.
\end{eqnarray}
Here, the ${\cal F}_\psi$ for $\psi=d,l$ are defined by
\begin{eqnarray}
{\cal{F}_\psi} \equiv {\cal{F}}(\tan^2 \theta_\psi, \tan^2\phi_\psi),
\end{eqnarray}
where 
\begin{eqnarray}
{\cal F}(a , b)&=&\frac{(ab)^{\frac{1}{4}}}{6(1-ab)^4(1+a)^{\frac{3}{2}}
(1+b)^{\frac{3}{2}}}
\left\{ 2(a+b)(-1+8ab-8 a^3 b^3 + a^4 b^4 +12 a^2 b^2 \ln (ab))
\right.
\nonumber \\
&& \left. -1-ab -64 a^2 b^2 +64 a^3 b^3 + a^4 b^4 + a^5 b^5
-36 a^2 b^2 (1+ab) \ln (ab) 
\right\}.
%
%
\end{eqnarray}
This function ${\cal F}(a , b)$ has the maximal value 0.1 at $a\simeq3$ and
$b\simeq3$. Eqs.(\ref{g1_mass})-(\ref{g3_mass}) imply that 
the so-called GUT relation of the gaugino masses does not
hold in general.

\newpage

%
%
%
\newcommand{\Journal}[4]{{\sl #1} {\bf #2} {(#3)} {#4}}
\newcommand{\APJ}{Ap. J.}
\newcommand{\CJP}{Can. J. Phys.}
\newcommand{\NC}{Nuovo Cimento}
\newcommand{\NP}{Nucl. Phys.}
\newcommand{\PL}{Phys. Lett.}
\newcommand{\PR}{Phys. Rev.}
\newcommand{\PRep}{Phys. Rep.}
\newcommand{\PRL}{Phys. Rev. Lett.}
\newcommand{\PTP}{Prog. Theor. Phys.}
\newcommand{\SJNP}{Sov. J. Nucl. Phys.}
\newcommand{\ZP}{Z. Phys.}

\clearpage
\begin{table}
\begin{center}
\begin{tabular}{|c|c|c|}  \hline 
 & &  \\
 & $Q, \psi, X$ & $Z_i$, $\psi'$
\\ \hline  
U(1)$_R$ & $0$ & $2$  \\ \hline
U(1)$_\chi$ & $1$ & $-2$ \\ \hline
\end{tabular}
\end{center}
\caption{U(1)$_R\times$ U(1)$_\chi$ charges for chiral superfields.
Here, $\psi=d,l,{\bar d},{\bar l}$ and $i=1,2,3$.}
\label{table_charge}
\end{table}
\clearpage
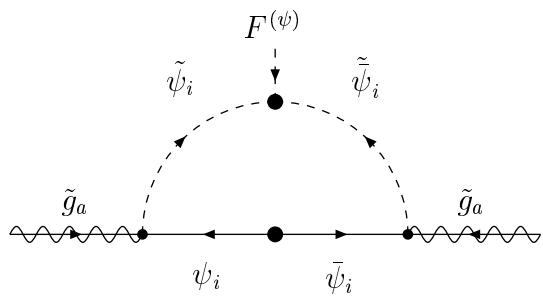
\begin{figure}
\caption{Diagram contributing to the gaugino masses 
where the single soft SUSY-breaking mass $F^{(\psi)}$ is inserted.
This contribution vanishes as shown in the Appendix.}
\label{gaugino_mass}
\end{figure}
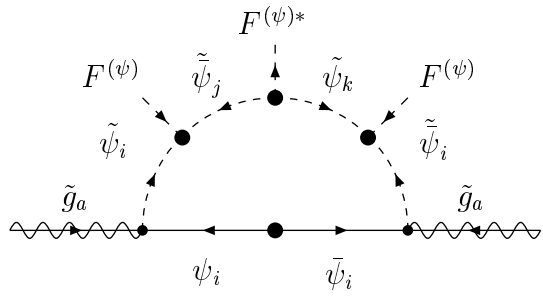
\begin{figure}
\begin{center} 
\caption{Diagram contributing to the gaugino masses 
where the three $F^{(\psi)}$'s are inserted.}
\label{3F_gaugino_mass}
\end{center}
\end{figure}
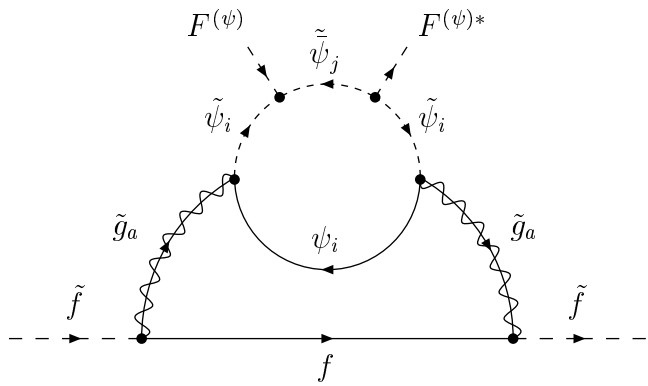
\begin{figure}
\caption{Typical two-loop diagram contributing to the sfermion masses.}
\label{two_loop_sfermion_mass}
\end{figure}
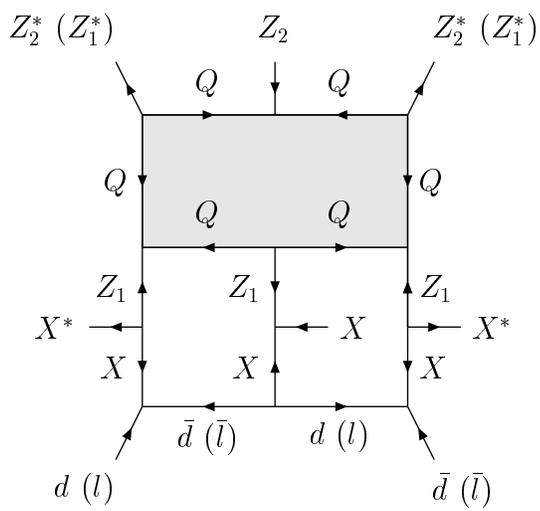
\begin{figure}
\caption{Typical diagram generating the effective K\"ahler potential
which contributes to the soft SUSY-breaking masses of the messenger
squarks and sleptons.}
\label{Kahler_corr}
\end{figure}
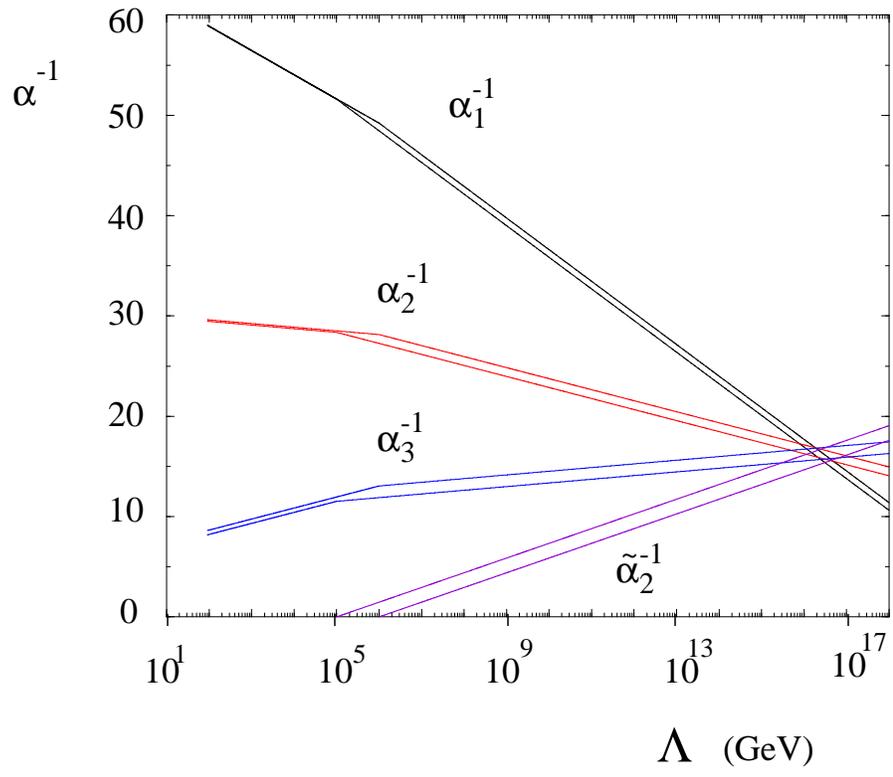
\begin{figure}
\caption{Renormalization group flow of the coupling constants
of SU(3)$_C$, SU(2)$_L$, U(1)$_Y$, and the strong SU(2) gauge groups.
Here, the mass of messenger squarks and sleptons is taken to be
$(10^5-10^6)~\GEV$ and we assume that the gauge coupling constant
$\tilde{\alpha}_2$ of the strong SU(2) becomes strong 
($\tilde{\alpha}_2/\pi \simeq 1$) at the scale $\Lambda= (10^5-10^6)~\GEV$.}
\label{gauge_coupling}
\end{figure}
\end{document}